\documentclass[manuscript]{aastex}

\newcommand{\gray}{$\gamma$-ray\ } \newcommand{\grays}{$\gamma$-rays\ }
\newcommand{\mic}{$\mu$m\ } 
\newcommand{\etal}{et al.\ }

\newcommand{\eg}{e.g.}

\begin{document}

\title{An Empirical Determination of the  Intergalactic Background Light Using NIR Deep Galaxy Survey Data out to 5 $\mu$m and the Gamma-ray Opacity of the Universe}

\author{Sean T. Scully}  
\affil{Department  of Physics and Astronomy,  James  Madison
University} 
\authoraddr{Harrisonburg, VA 22807; scullyst@jmu.edu}\authoraddr{Los  Angeles,  CA90095-1547}  
\author{Matthew A. Malkan}  
\affil{Department  of Physics  and  Astronomy, University  of
California,   Los  Angeles}  
\authoraddr{Los   Angeles,  CA90095-1547;
malkan@astro.ucla.edu}
\authoraddr{Harrisonburg, VA 22807; scullyst@jmu.edu}
\author{Floyd W. Stecker}  
\affil{Astrophysics Science Division, NASA/Goddard  Space   Flight  Center}
\authoraddr{Greenbelt, MD 20771; Floyd.W.Stecker@nasa.gov}
\affil{Department of Physics  and Astronomy, University of California,
Los  Angeles}

\begin{abstract}

We extend our previous model-independent determination of the intergalactic background light, based purely on galaxy survey data, out to a wavelength of 5 $\mu$m. Our approach enables us to constrain the range of photon densities, based on the uncertainties from observationally determined luminosity densities and colors.  We further determine a  68\% confidence upper and lower limit on the opacity of the universe to \grays up to energies of $1.6/(1+z)$ TeV. A comparison of our lower limit redshift-dependent opacity curves to the opacity limits derived from the results of both ground-based air \v{C}erenkov telescope and {\it Fermi} LAT observations of PKS 1424+240 allows us to place a new upper limit on the redshift of this source, independent of IBL modeling.

\end{abstract}

\subjectheadings{diffuse radiation -- galaxies: observations -- gamma-rays: theory -- galaxies: PKS 1424+240}

% Please see the AASTeX manual for a more complete discussion on how to make
% \cite-\reference work for you.   

\section{Introduction}

Past work on estimating the spectral and redshift characteristics of the intergalactic 
photon density (IBL) (with the $z = 0$ IBL usually referred to as the EBL), has depended on various assumptions as to the evolution of stellar populations and dust absorption in galaxies. There have also been attempts to probe the EBL using studies of blazar $\gamma$-ray spectra (Ackermann et al. 2012; Abramowski et al. 2013), an approach originally suggested by Stecker, De Jager \& Salamon (1992).  

We previously have pursued an independent, {\it fully empirical} approach to calculating the IBL deep using galaxy survey data (Stecker, Malkan \& Scully~2012 -- hereafter SMS; see also Helgason \& Kashlinsky 2012). In this paper we extend the previous SMS results on the $\gamma$-ray opacity of the Universe to TeV energies accessible to ground-based air \v{C}erenkov telescopes, here including galaxy survey data extending beyond the photometric $I$ band to the $M$ band ({\it i.e.}, 0.8 \mic to $\sim$ 5 \mic). We stress that the SMS approach, being completely observationally based, is {\it model independent}, relying only on published luminosity densities. Such an approach has a particular advantage over model-based methods, as it enables a determination of both the IBL {\it and its observational uncertainties} without making any assumptions about how the galaxy luminosity functions (LFs) evolve, as some approaches require ({\it e.g.} Malkan \& Stecker 1998, 
1; Kneiske, Mannheim \& Hartmann, 2002: Stecker, Malkan \& Scully 2006; Franceschini et al. 2008; Finke, Razzaque, \& Dermer, 2010; Kneiske \& Dole 2010) or by using semi-analytic models that make assumptions concerning galaxy evolution, stellar population synthesis models, star formation rates, and dust attenuation ({\it e.g.}, Gilmore et al. 2009; Somerville et al. 2011). Dom\'{i}nguez et al. (2011) used the redshift evolution of the $K$-band LFs of Cirasuolo et al. (2010), together with model templates based on {\it AEGIS} data to model the IBL. A recent detailed review of the problem has been given by Dwek \& Krennrich (2013). 

Direct integration of the galaxy survey data in the wavelength range from the FUV to $\sim$ 5 $\mu$m has now become possible. Observational data in this wavelength range, even out to redshifts $2 < z < 4$, have become sufficiently complete to eliminate the necessity of using model-based approaches. The high-redshift galaxies are sufficiently sampled to enable us to interpolate between observationally determined luminosity densities obtained for specific wavebands.  We use observationally determined galaxy colors and their associated errors to fill in the gaps in wavelength bands where there are less data.

Here we apply the technique of SMS to determine the IBL to a wavelength of $\sim$ 5 $\mu$m, thus extending our previous range of \gray opacities of the Universe to energies up to 1.6 TeV. This range is not only relevant for the sources with the highest redshifts observed by the {\it Fermi}-LAT telescope, but also for results in the sub-TeV and TeV energy range made using ground-based air \v{C}erenkov telescopes. Our technique to determine the expected \gray opacity of the Universe is thus complementary to the technique of using \gray observations directly, because the intrinsic (unabsorbed) spectra of the \gray sources are uncertain.

To further show the utility of our approach, we apply our lower limit opacities to the results of both ground-based air \v{C}erenkov telescope observations and {\it Fermi}-LAT observations of the BL Lac object PKS 1424+240, which allows us to place a {\it true upper limit} on the redshift of this source, independent of IBL models.  Previous observations of the ultraviolet absorption from the IGM of this object place a {\it lower limit} on its redshift of $z \ge .6035$ (Furniss \etal 2013).     

Our results give the \gray opacity as a function of energy and redshift to within an observationally determined 68\% confidence band. Therefore, a direct comparison with \gray spectral data will allow an independent determination of the previously suggested effects on such spectra. These effects include secondary \gray production generated by: (1) cosmic-ray interactions along the line of sight to the source (Essey \etal 2010; Essey \& Kusenko 2012) and (2)  line-of-sight photon-axion oscillations during propagation ({\it e.g.}, De Angelis et al. 2007). Both of these processes can produce an apparent reduction of the pair-production opacity effect as derived from $\gamma$-ray spectra alone. Thus our results can be critical in analyzing the implications of present and future $\gamma$-ray spectral data.

\section{Determination of the IBL from Galaxy Photon Emissivities}

Galaxy luminosity functions (LFs), $\Phi_{\nu}(L)$, can be determined by properly sampling galaxies in a survey for a given wavelength and accounting for biases.  As in SMS, in this work we chose to use only those references that give values for the integrated LF, {\it i.e.}, the luminosity density (LD), $\rho_{L_{\nu}}(z)$. This is because there is generally a lack of knowledge about the covariance of the errors in the Schechter function parameters used to determine the dominant statistical errors in their analyses.  Sufficient numbers of galaxies need to be observed up and down the luminosity function for a reliable constraint on their total luminosity density, {\it i.e.}, the integral of the LF.  It is not even guaranteed that a Schechter function will always provide the best analytic fit to the LF data.  The co-moving radiation energy density, $u_{\nu}(z)$, is derived from the equivalent co-moving specific emissivity ${\cal E}_{\nu}(z)$ = $\rho_{L_{\nu}}(z)$. 

\subsection{Luminosity Densities}

In this work, we use the same set of galaxy survey data and resulting luminosity
densities that covered rest-frame wavelengths from the far-ultraviolet to the $I$ band as in SMS; however we now extend our calculations to the $J, {\rm and} ~K$ bands.  We have excluded the $H$ band due to a paucity of observational data but we have checked that our approach is consistent with what data exists in this band.  We further use the colors derived in the next subsection to extend the calculation into the $L$ and $M$ bands (out to 4.8 $\mu$m).

The co-moving radiation energy density $u_{\nu}(z)$ is the time integral of the co-moving specific emissivity ${\cal E}_{\nu}(z)$,
\begin{equation} 
\label{u1}
u_{\nu}(z)=
\int_{z}^{z_{\rm max}}dz^{\prime}\,{\cal E}_{\nu^{\prime}}(z^{\prime})
\frac{dt}{dz}(z^{\prime})%e^{-\tau_{\rm eff}(\nu,z,z^{\prime})},
\end{equation}

\noindent where $\nu^{\prime}=\nu(1+z^{\prime})/(1+z)$ and $z_{\rm max}$ is the
redshift corresponding to initial galaxy formation 
(Salamon \& Stecker 1998), and
\begin{equation}
\frac{dt}{dz}{(z)} = {[H_{0}(1+z)\sqrt{\Omega_{\Lambda} + \Omega_{m}(1+z)^3}}]^{-1},
\label{cosmology}
\end{equation}

\noindent with $\Omega_{\Lambda} = 0.72$ and $\Omega_{m} = 0.28$.
We note that we have scaled all of the observational data for ${\cal E}_{\nu}(z)$ to 
a value of h = 0.7 for consistency. %We can therefore write equation (\ref{u1}) as

%\begin{equation} 
%\label{u2}
%u_{\nu}(z)=
%\int_{z}^{z_{\rm max}}dz^{\prime}\,{\cal E}_{\nu^{\prime}}(z^{\prime})
%\frac{dt}{dz}(z^{\prime}){{\cal H}(\nu(z') - \nu'_{LyL})},
%\end{equation}

%\noindent where ${\cal H}(x) $ is the Heavyside step function.

\subsection{Average Colors}

The continuum emission from galaxies between 0.8 $\mu$m and 5 $\mu$m arises predominantly from stellar photospheres. At these near infrared (NIR) wavelengths, the light is mostly emitted by red giant stars.  Within a few million years of the first generation of star formation, massive stars will have left the main sequence to begin populating the red giant branch.  At later times, red giant branch stars do not vary greatly from one stellar population to another. The resulting spectral energy distributions (SEDs) of galaxies are therefore much more similar to each other in the NIR range than they are at shorter wavelengths. The scatter in red and NIR SEDs of galaxies is especially small, since all of them contain old, red giant branch stellar populations, which dominate the stellar mass and the continuum emission around 1 $\mu$m.

Dai \etal (2009) determine galaxy luminosity functions at 3.6, 4.5, 5.8, and 8.0 \mic for a sample of 3800 -- 5800 galaxies utilizing combined photometry from the {\it Spitzer}/IRAC Shallow Survey with redshifts from the AGN and Galaxy Evolution Survey of the NOAO Deep Wide-Field Survey Bo\"otes field. They obtained well defined complete luminosity functions in the local redshift bin of $z \le$ 0.2. They then derived galaxy luminosity densities and generated a best-fit SED (their figure 13).  We take these data to be representative of low redshift galaxies ($z \le$ 0.5).

To estimate the red and NIR SEDs of galaxies at higher redshifts, we have used the template SEDs derived by Kriek \etal (2010,2011)\footnote{see also {\tt http://astro.berkeley.edu/~mariska/comp/}}. Those authors utilized extensive multi-band photometry obtained for a sample of 3500 $K$-band selected galaxies, at redshifts between $z = 0.5$ and $z = 2.0$, close to a mass-limited sample of galaxies.  For redshifts between $z = 0.5$ and $z = 1.0$, the galaxy stellar masses are uniformly distributed between the range of 10$^{9.7}$ and 10$^{11.3}$ M$_\sun$.  At redshifts from 1 to 2, the mass range covered by these SEDs shrinks to 10$^{10.3}$ to 10$^{11.3}$.
Kriek \etal (2010) grouped their galaxy SEDs into 30 average templates spanning the full range from the bluest to the reddest galaxies.   
These typical $L_{*}$ galaxies are the ones that produce most of the cosmic red and NIR emission. Therefore we take the unweighted average of the Kriek \etal (2010,2011) SEDs to represent the full sample of galaxies.
This should be reasonably accurate as there is only a 0.5 magnitude $Y - L$ color variation  (from 1.05 to 3.5 $\mu$m) between the bluest and reddest galaxies.  Therefore, over this full wavelength range, the {\it average} colors we have adopted must be quite close to the colors of any galaxy, at least at these redshifts.  We express the average colors we have adopted as flux ratios normalized to $J$ band in Table 1.
\begin{deluxetable}{ccccc}
\tabletypesize{\footnotesize}
\tablecolumns{5}
\tablewidth{0pt}
\tablecaption{ Average Colors \label{table:colors}}
\tablehead{
\colhead{Waveband} & \colhead{Relative $\nu L_\nu$}   &  \colhead{Std. Deviation} & \colhead{Min Value}
&  \colhead{Max Value}}
\startdata
$J$ (1.2 $\mu$m) & 1$^*$ & -- & -- & -- \\ 
$H$ (1.6 $\mu$m) & 1.19 & 0.08 & 1.09 & 1.37 \\
$K$ (2.2 $\mu$m) & 0.97 & 0.12 & 0.83 & 1.23 \\
$L$ (3.5 $\mu$m) & 0.70 & 0.15 & 0.43 & 0.99 \\
$M$ (4.8 $\mu$m) & 0.50 & 0.16 & 0.3 & 0.84
\enddata
\vspace{-0.8cm}
\tablecomments{The bias in the colors towards
bluer averages occurs because there are not many galaxies in the very red categories, so the mean color stays on the blue side of (Max + Min)/2. $^*$All differences are relative to the $J$ band.}
\end{deluxetable}

The peak in the galaxy SEDs that we used occurs around a rest wavelength of 1.6 $\mu$m. We note that there is a subtle trend for the red and NIR SEDs to become bluer at higher redshift. This is because the younger stellar populations were more important in the early universe.  This effect produces a small shift of the SED peak to slightly shorter rest wavelengths at higher redshifts.  This trend may continue to redshifts higher than 2. However, because this effect is small, we assume that the SEDs of Kriek et al. (2010) apply to all higher redshifts.   We assume that these average colors also apply to the less massive galaxies that Kriek et al. could
not study.  Fortunately, the total NIR emission of galaxies tends to be dominated by the more massive galaxies (\eg., Ly et al 2011).  Furthermore, galaxies of masses smaller by an order of magnitude are only slightly bluer.

\subsection{Photon Density Calculations}

As in SMS, we derive a luminosity confidence band in each of our additional wavebands by using a robust rational fitting function characterized by
\begin{equation}
\rho_{L_{\nu}} = {\cal E}_{\nu}(z) = {{ax+b}\over{cx^{2}+dx+e}}
\label{rational}
\end{equation}
where $x = \log(1+z)$ and $a$, $b$, $c$, $d$, and $e$ are free parameters.  We compute the 68\% confidence band from Monte Carlo simulations by finding 100,000 realizations of the data and then fitting the rational function given by equation (\ref{rational}).  As in SMS, we model symmetric error bars with a Gaussian distribution, while choosing a skew normal distribution to model asymmetric errors.  

The $K$ band data do not extend beyond a redshift of $z$ $\sim$ 2, while the $J$ band data extend to $z\sim$ 3. Thus, the fits determined using the above method can not be trusted beyond these limits.  To handle this issue, we truncate our fits at a redshift of $z = 2$, and then color transform our previous $I$ band from SMS to fit at the higher redshifts. This is a reasonable assumption owing to the expected similarity in the SEDs of galaxies at these wavelengths (see previous sections). Indeed, the overlapping $J$ band data beyond a redshift of 2 are in good agreement with this color-shifted band. 

Figure 1 shows the redshift evolution of the luminosity density for the various wavebands based on the survey data published in the literature for the wavelength bands that we use to extend the previous work of SMS.\footnote{References for the values of ${\cal E}_{\nu}(z)$ used to construct Figure 1 are listed in the legend.} The upper and lower limits of the bands correspond to the highest and lowest IBL consistent with 68\% confidence errors. The error bars on the $J$ and $K$ band data taken from Pozzetti \etal (2003) are small in comparison to other data used. Their uncertainty was estimated strictly by considering the range of acceptable Schechter parameters when fitting their LFs. As other authors considered  systematic errors such as cosmic variance and extrapolation to the faint end, we have chosen to assign a linear error of 20\% to these data to bring their errors more in line with what is typically quoted so that these data do not inordinately influence the fitting routine. The upper and lower limits of the bands correspond to the highest and lowest IBL consistent with 68\% confidence errors.

Our method of dealing with the confidence band for redshifts beyond where there are any data for $J$ and $K$ is not critical for the opacity calculation. There is very little contribution to the \gray opacities from photons in these wavebands beyond a redshift of 2 because of the short time interval of the emission from galaxies at higher redshifts (see equation (\ref{cosmology})). To verify this, we set an upper limit by assuming that the value of the photon density at $z = 2$ remains constant out to higher redshifts. Since we expect that the LD
will drop off at the higher redshifts, this assumption gives an upper limit. We found no appreciable differences in the calculated opacities under this assumption (see next section) as expected.  

Having determined the colors between the $J, K, L, {\rm and} ~M$ bands, we use them to transform the $J$ and $K$ LDs into the $L$ and $M$ bands and again apply our fitting technique to produce confidence limits for those wavelengths.  The color-transformed data along with the confidence bands for $L$ and $M$ can also be seen in Figure 1.  We note that the wider resulting confidence bands for $L$ and $M$ occur because we have propagated the error determined from the colors to the $J$ and $K$ data (reflected in the size of the error bars) before running the Monte Carlo simulations and fits. This is as expected, since the band of acceptable photon densities derived from the data should reflect both errors. As in SMS, we then interpolate between the upper limit and lower limit of the confidence bands between the various wavebands separately to find the upper and lower limit rest frame luminosity densities. 

\section{Determination of the Optical Depth to \grays and resulting \gray Horizon} 

Using the results of the previous section, the specific emissivity can then be derived for the highest and lowest IBL LDs. The co-moving radiation energy density is then determined from equation (\ref{u1}). The photon densities derived thereby are given by
\begin{equation}
n(\epsilon,z) = u(\epsilon,z)/\epsilon \ \ ,
\label{gammadens}
\end{equation}
with $\epsilon = h\nu$, and
with $u(\epsilon,z)$ given by equation (\ref{u1}).

With the co-moving photon density $n(\epsilon,z)$ evaluated, the optical
depth for \grays owing to electron-positron pair production 
interactions with photons of the stellar radiation
background can be determined from the expression (Stecker, De Jager \& Salamon ~1992)
\begin{equation} \label{G}
\tau(E_{0},z_{e})=c\int_{0}^{z_{e}}dz\,\frac{dt}{dz}\int_{0}^{2}
dx\,\frac{x}{2}\int_{\epsilon_{th}}^{\infty}d\epsilon\,(1+z)^{3}n(\epsilon,z)
\sigma_{\gamma\gamma}[s(z)],
\label{tau}
\end{equation}
In equation 
(\ref{tau}), $dt/dz$ is given by equation (\ref{cosmology}), $E_{0}$ is the observed \gray energy at redshift zero, 
$\epsilon$ is evaluated at
redshift $z$,
$z_{e}$ is the redshift of
the \gray source, $x=(1-\cos\theta)$, $\theta$ being the angle between
the \gray and the soft background photon, and
the pair production cross section $\sigma_{\gamma\gamma}$ is zero for the rest system
center-of-mass energy $\sqrt{s} < 2m_{e}c^{2}$, $m_{e}$ being the electron
mass.  Above this threshold, the pair production cross section is given by
\begin{equation} \label{H}
\sigma_{\gamma\gamma}(s)=\frac{3}{16}\sigma_{\rm T}(1-\beta^{2})
\left[ 2\beta(\beta^{2}-2)+(3-\beta^{4})\ln\left(\frac{1+\beta}{1-\beta}
\right)\right],
\end{equation} 
where $\sigma_T$ is the Thompson scattering cross section and $\beta=(1-4m_{e}^{2}c^{4}/s)^{1/2}$  (Breit \& Wheeler 1934; Jauch \& Rohrlich 1955).

As derived in SMS, the pair-production cross section energy has a threshold at $\lambda = 4.8 \ \mu {\rm m} \cdot E_{\gamma}({\rm TeV})$, determined from the energy required to produce twice the electron rest mass in the center of mass frame. Since the maximum $\lambda$ 
is in the rest frame M band at 4.8 $\mu$m at redshift $z$, the resulting energy at
threshold is $\sim$ 1.6 TeV at $z = 0$.  The energy at
interaction in the rest frame is given by $(1+z)E_{\gamma}$ meaning the maximum \gray energy affected by the photon range that we consider is $\sim 1.6(1+z)^{-1}$ TeV.

The 68\% opacity ranges for $z = 0.1, 0.5, 1, 3 ~$and $5$, calculated using the SMS methods as described above, are plotted in Figure 2. The increasing uncertainties in the \gray opacity towards higher redshifts are a reflection of the increasing widths of the uncertainty bands in the luminosity densities shown in Figure 1, as follows from equations (\ref{gammadens}) and (\ref{tau}).  

Figure 3 shows  an energy-redshift plot of the highest energy photons from extragalactic sources at various redshifts from {\it Fermi} as given by Abdo et al. (2010) along with our 68\% confidence band for $\tau = 1$,  extending our result from SMS down to a redshift of $z = 0.2$. We stress that our 68\% confidence band as shown in Figure 3 is the range of optical depths allowed as derived from our observationally determined IBL results. 

\section{Application of our results to PKS 1424+240}

To show the utility of our results, we consider redshift limits obtained from \gray observations of the distant blazar PKS 1424+240. BL Lac objects such as this one typically display bright and featureless continunous spectra making definite spectroscopic determinations of their redshifts challenging. A recent determination of a {\it lower limit} of the redshift of PKS 1424+240 has been obtained by Furniss \etal (2013). They find $z \ge$ 0.6035, inferred from {\it Hubble} observations of the measured positions of Lyman lines from intervening Lyman $\alpha$ absorbers at lower redshifts. PKS 1424+240 has been observed at \gray energies lower than 100 GeV by {\it Fermi-LAT} and  at energies above 500 GeV by $VERITAS$ (Acciari \etal 2010).  

Using our opacity results we can place an {\it upper limit} on the redshift of this source.  A  power-law spectrum with index $ \sim 1.8$ has been determined from {\it Fermi} for energies below 100 GeV (Furniss \etal 2013). In this energy range there should be little \gray attenuation. Therefore, extending this power-law spectrum out to energies measured by $VERITAS$ should represent an upper limit on its intrinsic spectrum.
An upper limit for the opacity for each of the measured \gray energies can then be determined from the flux difference between the extended power-law spectrum and the measured fluxes.  This upper limit can then be compared with our redshift dependent opacity curves generated from our lower limit IBL, thereby determining an upper limit on the redshift of PKS 1424+240.  Our upper limit on $z$ is found to be 1.0; our lower limit opacities for redshifts greater than $z = 1.0$ exceed the upper limits on the opacity determined from the flux differences.  Figure 4 shows the upper limits on the opacities determined from the {\it VERITAS} data along with our lower limit $z = 1.0$ opacity curve. 
Aleksi\'{c} et al.~(2014) have used {\it MAGIC} data to place an upper limit on the redshift of PKS 1424+240 of 0.81. However, the limit obtained by the {\it MAGIC} collaboration depends on assuming absolute accuracy of the IBL model of Franceschini et al. (2008). The loosening of this limit to a slightly larger upper limit of $z = 1.0$ is a result of using our observationally derived uncertainty band rather than a theoretical model. 

\section{Comparison with Other Results}

Helgason \& Kashlinsky (2012) (HK) have used a method similar to ours, using galaxy LFs to determine the $\gamma$-ray opacity of the universe with uncertainties over the same energy range as considered here. Their results are also model independent. However, their treatment differs from our ours in two respects: (1) The HK uncertainty band is determined by fitting to Schechter function parameterizations with upper and lower limits determined by fitting to faint end slopes of the Schecter functions; we use observationally determined LDs with observational errors combined with our Monte Carlo treatment as described in Section 2.3. Our use of observationally determined errors takes account of systematic effects such as cosmic variance. (2) We use color data as described in Section 2.3 to extrapolate our LDs to higher redshifts, allowing us to make use of more observational data, particularly in the $L$ and $M$ bands; HK postulate an exponential cutoff in their LDs at higher redshifts. The result of these differences is that our opacities are somewhat larger than those of HK.

Our uncertainty band for the IBL at $z = 0$ (the EBL) is in full agreement with that obtained by Abramowski \etal (2013) using \gray spectra from {\it H.E.S.S.} (see their Figure 5). Our results are consistent with the work of the {\it Fermi} collaboration (2010) as shown in our Figure 3. Our results for $z = 1$ are in fair agreement with those of the analyses of the {\it Fermi} collaboration (Ackermann et al. 2012). Our $\tau = 1$ (horizon) band is consistent with the result of Dom\'{i}nguez \etal (2013).  

\section{Conclusions}

We have presented an extension of our previous determination of a 68\% confidence bands giving
upper and lower limits on the IBL out to 4.8 $\mu$m.  This model-independent determination is entirely based on observationally derived luminosity functions from local and deep galaxy survey data and color data. This has enabled us to directly derive both the $\gamma$-ray opacity and its
observational uncertainties as a function of both energy and redshift out to an energy of
$1.6/(1 + z)$ TeV for $z \le 5$. We have applied our lower limit opacities to the results of both ground-based air \v{C}erenkov telescope and {\it Fermi} LAT observations of PKS 1424+240, allowing us to place a new upper limit on the redshift of this source, {\it viz.}, $z \le 1.0$, independent of IBL modeling. 

We find no direct evidence in our spectral analysis for a required modification of our predicted $\gamma$-ray opacities as shown in Fig. 4, and for the opacity in the direction of PKS1424+240, either owing to axion-photon mixing as suggested Meyer \& Horns (2013) or by the existence of secondary production effects as suggested by Essey \& Kusenko (2013). However, such effects may need to be invoked should a subsequently determined redshift of PKS 1424+240 turn out to be greater than 1.

\section*{Acknowledgments}

We wish to thank Amy Furniss for supplying us with the {\it VERITAS} $\gamma$-ray data on PKS1424+240
that we used in our analysis. We also thank Mariska Kriek for supplying us with galaxy template 
SEDs.

%\clearpage

%\begin{thebibliography}{}

%\end{thebibliography}{}

\clearpage

\normalsize

\centerline{Figure Captions}

\noindent Figure 1: The specific emissivities for $J, K, L, {\rm and} ~M$ wavebands.  In the $L$ and $M$ panels, the $J$ and $K$ band data have been shifted using the color relations given in the text in order to fully determine the specific emissivities in these wavebands. The grey shading represents the 68\% confidence bands (see text).

\vspace{10pt}

\noindent Figure 2: The empirically determined opacities for redshifts of 0.1, 0.5, 1, 3, 5 extended from SMS. The dashed lines indicate the opacities $\tau = 1$ and $\tau = 3$.

\vspace{10pt}

\noindent Figure 3: A $\tau = 1$ energy-redshift plot (Fazio \& Stecker 1970) showing our uncertainty band results compared with the {\it Fermi} plot of their highest energy photons from FSRQs (red), BL Lacs (black) and and GRBs (blue) {\it vs.} redshift (from Abdo et al. 2010).

\vspace{10pt}

\noindent Figure 4: Determination of the upper limit redshift for PKS 1424+240.  The points represent the upper limits to the \gray opacity derived from the $VERITAS$ data.  The dashed curve corresponds to our lower limit opacity for $z = .6035$ while the  solid curve is our best fit lower limit opacity to the opacity limits derived from the results shown in Fig. 3, corresponding to an upper limit redshift of $z = 1.0$.

\clearpage

\begin{figure}
\begin{center}
\includegraphics[width=7in]{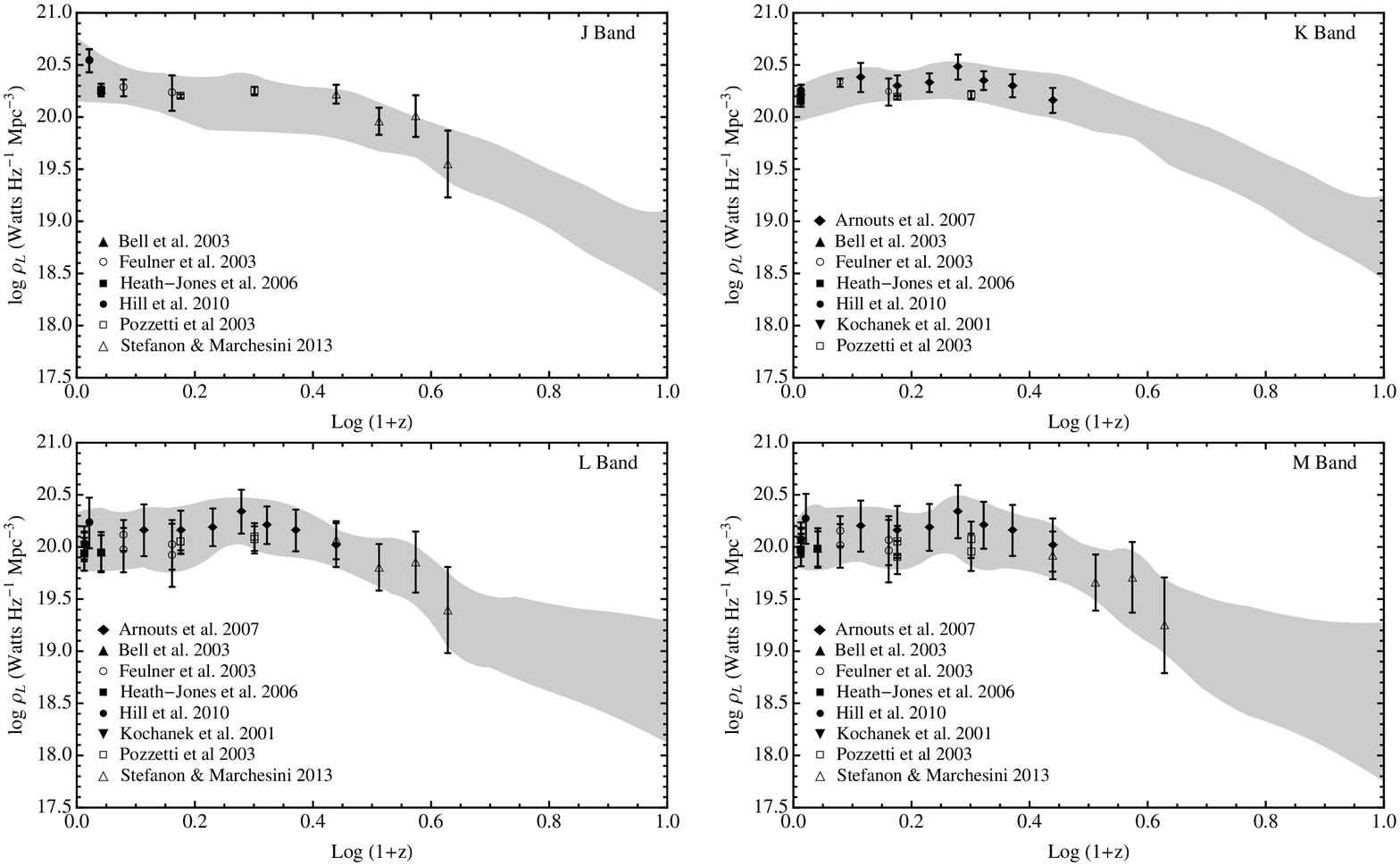}
Figure 1.
\label{confband}
\end{center}
\end{figure}

\clearpage

\begin{figure}
\begin{center}
\includegraphics[width=7in]{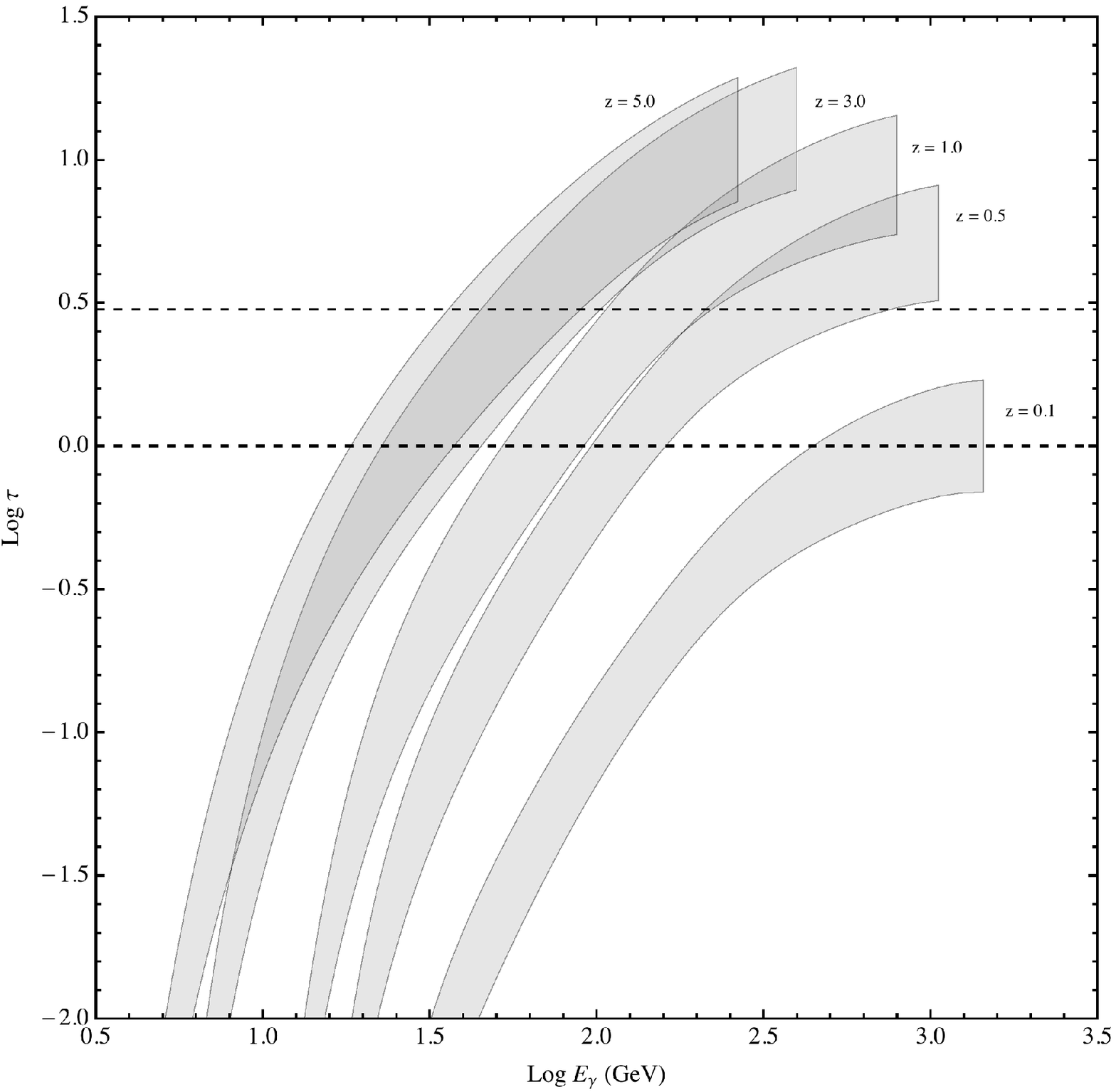}
Figure 2.
\label{opacities}
\end{center}
\end{figure}

\clearpage

\begin{figure}
\begin{center}
\includegraphics[width=7in]{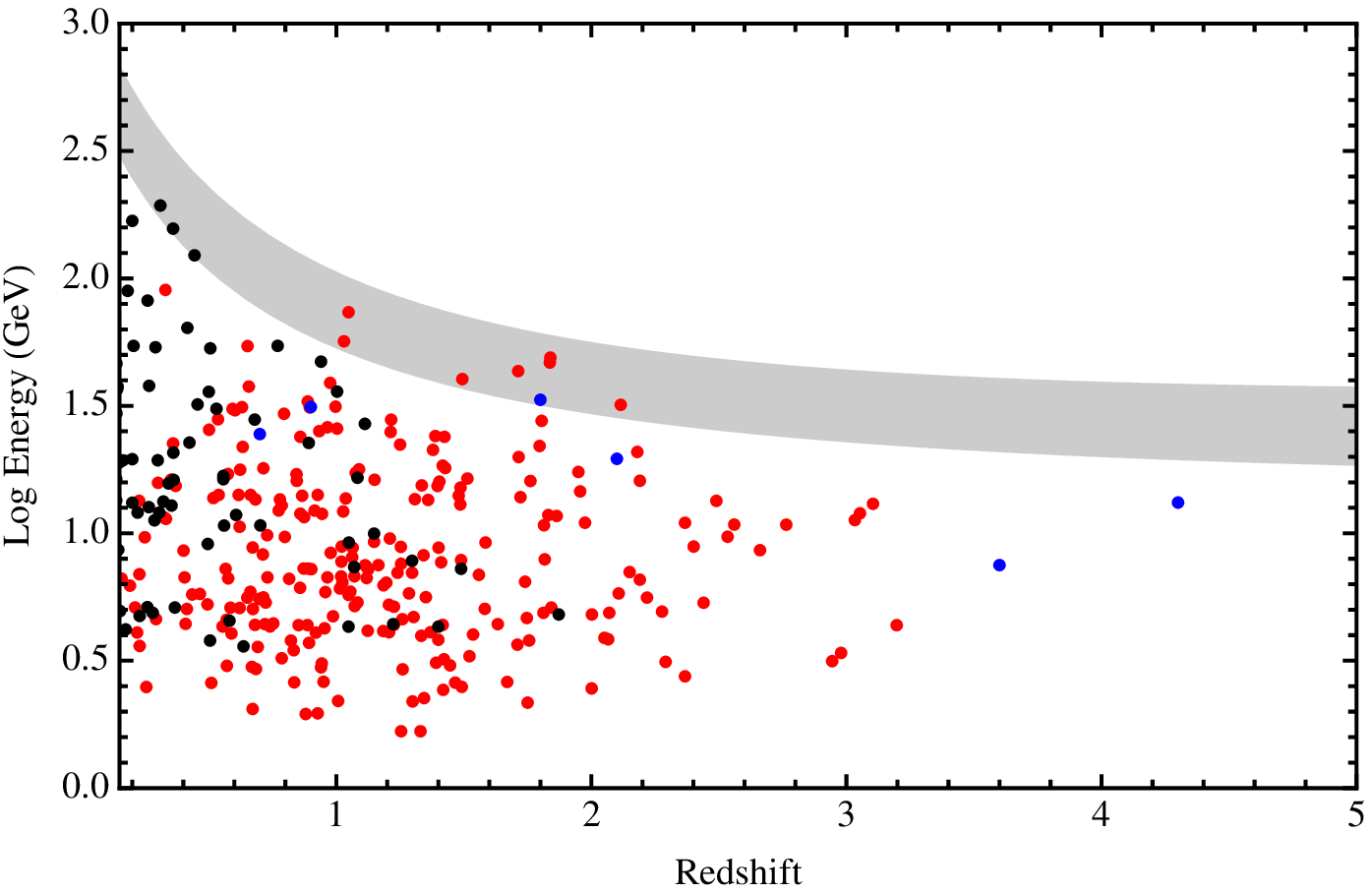}
Figure 3. 
\label{FSplot}
\end{center}
\end{figure}

\clearpage

\begin{figure}
\begin{center}
\includegraphics[width=7in]{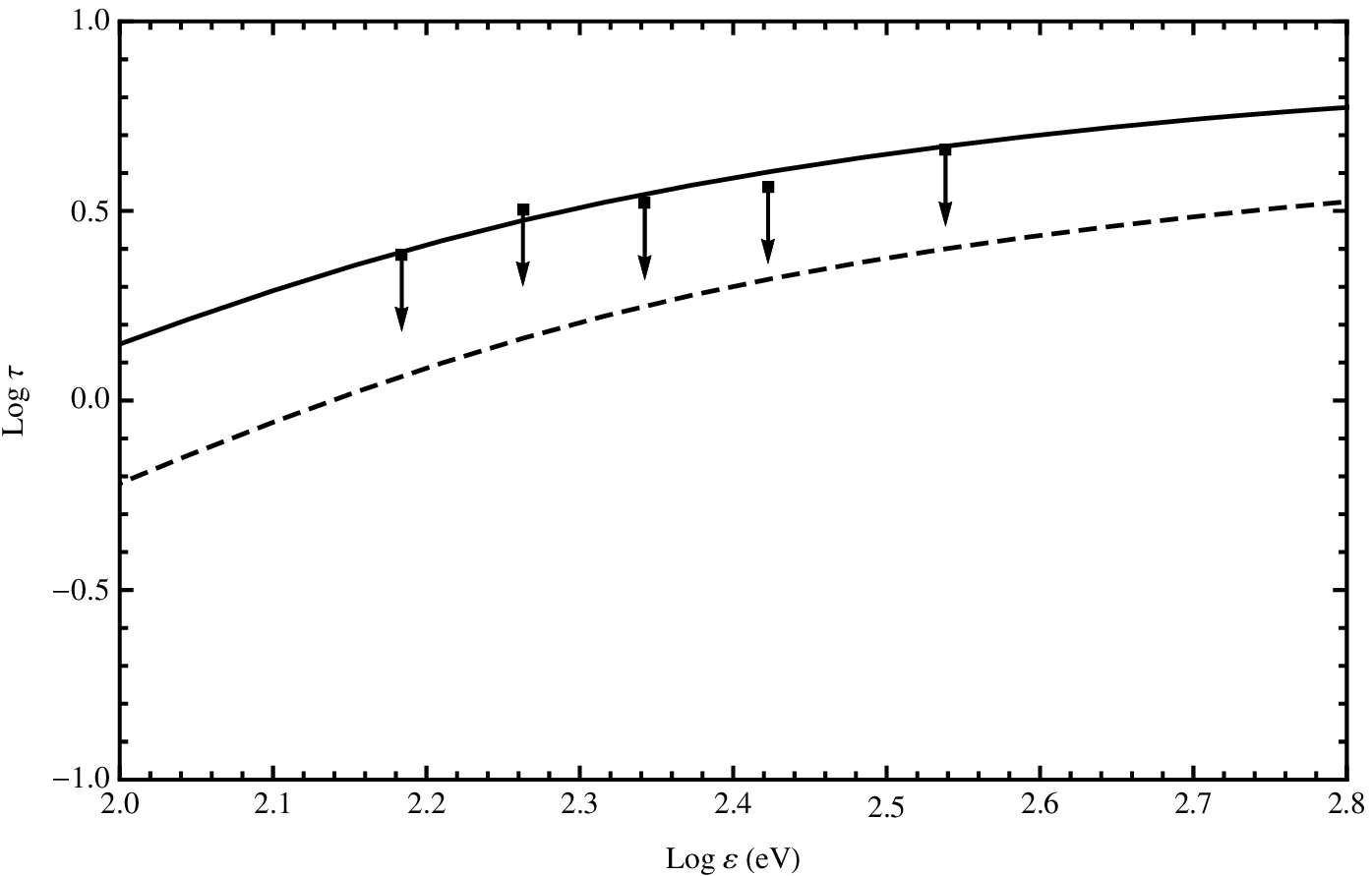}
Figure 4. 
\label{PKS1424}
\end{center}
\end{figure}

\end{document}